\documentclass[10pt,aps,showpacs,floatfix,twocolumn,amsmath,amssymb,groupedaddress,superscriptaddress]{revtex4}
\usepackage{epsfig}
\usepackage{graphicx}
\usepackage{graphics}
\usepackage{xspace}
\usepackage[dvips]{color}
\usepackage{latexsym}
\usepackage{natbib}
\usepackage{mathrsfs}
\newcommand{\inieq}{\begin{eqnarray}}            
\newcommand{\fineq}{\end{eqnarray}}            
\newcommand{\diff}{{\rm\,d}}                    
\newcommand{\bint}{\mskip .5mu \int \mskip-18mu} 
\newcommand{\be}{\begin{equation}}
\newcommand{\ee}{\end{equation}}
\newcommand{\ba}{\begin{eqnarray}}
\newcommand{\ea}{\end{eqnarray}}

\def\q{\mbox{\boldmath $q$}}

\def\k{\mbox{\boldmath $k$}}

\begin{document}
\title{Relativistic descriptions of quasielastic charged-current 
neutrino-nucleus scattering: application to scaling and superscaling ideas}
\author{Andrea Meucci} 
\affiliation{Dipartimento di Fisica Nucleare e Teorica, 
Universit\`{a} degli Studi di Pavia and \\
Istituto Nazionale di Fisica Nucleare, 
Sezione di Pavia, I-27100 Pavia, Italy}
\author{J.A. Caballero}
\affiliation{Departamento de F\'\i sica At\'omica, Molecular y
Nuclear, Universidad de Sevilla, E-41080 Sevilla, Spain}
\author{C. Giusti}
\affiliation{Dipartimento di Fisica Nucleare e Teorica, 
Universit\`{a} degli Studi di Pavia and \\
Istituto Nazionale di Fisica Nucleare, 
Sezione di Pavia, I-27100 Pavia, Italy}
\author{J.M. Ud\'{\i}as}
\affiliation{Grupo de F\'{\i}sica Nuclear, Departamento de F\'\i sica At\'omica, Molecular y
Nuclear, Universidad Complutense de Madrid, E-28040 Madrid, Spain}

\date{\today}

\begin{abstract}
The analysis of the recent experimental data on charged-current neutrino-nucleus 
scattering cross sections measured at MiniBooNE requires  relativistic theoretical 
descriptions also accounting for the role of final state interactions.
In this work we evaluate inclusive quasielastic differential neutrino cross sections 
within the framework of the relativistic impulse approximation. Results based on the 
relativistic mean field potential are compared with the ones corresponding to the 
relativistic Green function approach. An analysis of scaling and superscaling properties 
provided by both models is also presented.
\end{abstract}


\maketitle


\section{Introduction}
\label{intro}

From the analysis of the world $(e,e')$ data at quasielastic (QE) kinematics, i.e., 
with non nucleonic degrees of freedom playing a minor role,
it has clearly emerged the validity of scaling arguments for values of the transferred 
momentum $q$ high enough~\cite{Day90,don1,don2,mai1}. Furthermore, the study of the 
separated longitudinal/transverse (L/T) data has shown that whereas the L response 
superscales, scaling violations are visible in the T channel due to effects beyond
the impulse approximation (IA): correlations, meson exchange currents (MEC), etc. 
These results constitute strong constraints for any theoretical model applied to 
electron scattering reactions; not only scaling and superscaling behavior should 
be fulfilled, i.e., independence on $q$ and 
on the nuclear target, but also the model should be capable to reproduce the 
specific shape of the experimental superscaling longitudinal function that 
presents a significant tail extended to large values of the transferred energy, 
that is, positive values of the superscaling variable $\psi$ ~\cite{mai1}.

A large variety of theoretical models used in the literature satisfy scaling arguments. 
This is the case of the relativistic Fermi Gas (RFG) model that superscales by
construction~\cite{don1,don2,mai1}. Other approaches based on non-relativistic reductions 
also fulfill scaling behavior even in presence of final state interactions (FSI)
and nucleon correlations (see 
Refs.~\cite{amaro05,neutrino2,antonov06,antonov07,amaro07}).
However, most of them lead to scaling functions that do not get asymmetry, being in clear 
disagreement with data. Our past studies have shown that there are two basic aspects in 
the description of electron scattering which are necessary to compare superscaling 
properties of different models with experimental data: relativity and FSI. On the one 
hand, scaling comes out only when the momentum transfer is high enough. 
This means that, not only relativistic kinematics should be considered, but also the 
nuclear dynamics and the description of the operators should involve relativity. 
On the other hand, the importance of FSI has also been clearly stated in the case of 
exclusive $(e,e'N)$ processes where the use of complex optical potentials is 
required~\cite{Chinn89,Ud1,Ud3,Ud4,meucci1,meucci2,meucci3,Udi-PRL,Cris04}. However, even being 
restricted to the QE kinematical domain, the analysis of inclusive reactions, contrary 
to exclusive
ones, needs all inelastic channels in the final state to be retained, i.e., the flux 
should be conserved. Therefore, the distorted wave impulse approximation based on the 
use of a complex potential should be dismissed for $(e,e')$ processes.

Previous studies~\cite{cab2,cab1,isospin07,confee} have clearly illustrated the key role 
played by FSI to
reproduce the specific asymmetric shape shown by the experimental scaling function. 
Specifically, description of the final nucleon state using the same relativistic mean
field potential considered in describing the initial state, denoted simply as RMF, leads 
to a superscaling function in accordance with data. Recently~\cite{confee}, we
have extended this analysis by introducing the relativistic Green function (RGF) 
technique developed by the Pavia group~\cite{ee,cc,eenr,eeann}. This is an alternative,
relativistic, approach to FSI in $(e,e')$ reactions. Both descriptions, 
RMF and RGF, contrary to previous models based on non-relativistic or semirelativistic
reductions~\cite{amaro05}, provide a significant asymmetry in the scaling function 
following the general behavior of data. It is also important to point out that 
asymmetry in the
scaling function can be also obtained within the framework of a semirelativistic 
(SR) model, provided that FSI are described by the Dirac equation-based (DEB) potential 
derived
from the RMF~\cite{amaro07}.

High-quality predictions for neutrino-nucleus cross sections are needed for use in 
on-going experimental studies of neutrino oscillations at GeV energies. Recently,
muon charged-current neutrino-nucleus double differential cross 
sections have been measured for
the first time by the MiniBooNE collaboration~\cite{miniboone}. 
  The experimental conditions have made it possible to disentangle data for which 
no pions in the final state are detected. This corresponds to the QE regime where
only nucleons are the relevant degrees of freedom, and is simply denoted as 
charged-current quasielastic (CCQE) neutrino-nucleus scattering. 
The analysis of CCQE data shows that the RFG model 
understimates the total cross section, unless the axial mass $M_A$ is 
significantly enlarged with respect to the accepted world average value. The kinematics 
involved in the MiniBooNe experiment: neutrino energy flux ranging from 375 MeV up to 
$\sim 3$ GeV, requires the use of a 
relativistic description of the process. Moreover, a proper analysis of data requires also 
to have good control on nuclear effects
\cite{cab1,cab2,amaro05,cc,Cris06,Lava,Meucci:2006ir,Meucci:2008zz,
Kim:1994zea,Volpe:2000zn,Nieves:2005rq,
Butkevich:2007gm,Butkevich:2010cr,Benhar:2005dj,
Benhar:2006nr,bleve,Martini:2009uj,Leitner:2006ww,Buss:2007ar,Leitner:2008ue}. 
Notice that any reliable calculation for neutrino
scattering should first be tested against electron scattering in the same kinematical 
conditions. In this sense, while the RFG is capable of getting the basic size and shape of 
the inclusive cross section, it can hardly account for important details of the 
response. Thus, the need for a modification of the axial mass within 
the RFG could be a way to effectively incorporate nuclear effects into the 
calculation. 
As mentioned, relativity is expected to play an important
role at MiniBoone conditions. Not only relativistic kinematics, already included
in the RFG model, but also dynamical relativistic effects, not present in RFG,  
can be important. In this work, the two QE-based approaches selected, RMF and RGF, 
incorporate relativistic dynamics by means of strong
scalar and vector potentials in the Dirac equation employed to describe nucleons in nuclei.

As shown in Refs.~\cite{cab2,cab1,isospin07,confee}, the RMF and 
RGF approaches describe successfully 
the behavior of electron scattering data and their related scaling and superscaling 
 functions. Therefore, in this work we extend our previous analysis
for $(e,e')$ reactions to charged-current (CC) neutrino-nucleus scattering. The 
differences observed between the predictions of the two approaches can be helpful for 
understanding nuclear effects, particularly FSI, which may play a crucial role in the 
analysis of neutrino-nucleus scattering data and its influence in studies of neutrino 
oscillations at intermediate to high energies.

The paper is organized as follows. In Section II we summarize the basic formalism involved 
in the description of CC neutrino-nucleus scattering. We introduce the main expressions 
and present a brief discussion on the two approaches considered in the description of 
final state interactions, namely, the relativistic mean field and the relativistic Green 
function technique. Scaling and superscaling arguments 
for neutrino-nucleus scattering processes are also briefly mentioned. In Section III we 
show and discuss the results obtained with the two models considered. After checking the 
reliability of both calculations in the relativistic plane wave approach and also making 
use of the real part of the relativistic optical potential, we describe FSI within the RMF 
and RGF models, and present results for the differential cross sections and scaling 
functions. Finally, in Section IV we summarize our conclusions.


\section{Neutrino-nucleus scattering}
\label{sec.for}

The single-nucleon cross section for an inclusive reaction where an incident neutrino 
(antineutrino), with four-momentum $k^\mu = (E_{\nu},\k)$, is absorbed by a nucleus and 
only the outgoing muon, with four-momentum 
$k^{\prime\mu} = (\varepsilon^{\prime},\k^{\prime})$ and mass 
$m \simeq 105.7$ MeV is detected, is given by~\cite{amaro05,neutrino2,Yasuo95,Walecka} 
\begin{eqnarray} \label{eq.sigma}
	\frac{\diff \sigma}{\diff \Omega ^{\prime} \diff \varepsilon ^{\prime}} = 
	\sigma _0\ {\cal F}^2_{+(-)} \ ,
\end{eqnarray}
where
\begin{eqnarray}
	\sigma _0 &=& \frac{\left( G\cos\vartheta_C \right )^2}{2\pi ^2} k^{\prime}
	\varepsilon^{\prime}\cos^2\left(\overline{\vartheta}/2\right) \ , 
	\label{eq.sigma0} \end{eqnarray}
$G \simeq 1.166 \times 10 ^{-5}$ GeV$^{-2}$ is the Fermi constant, $\vartheta_C$ is 
the Cabibbo angle with $\cos^{2}\left( \vartheta_C\right) \simeq 0.975$, and  
$\overline{\vartheta}$ is the generalized scattering angle, that is defined by
\begin{equation}
	\tan^2\left(\overline{\vartheta}/2\right) = \frac{|Q^2|}{\left(E_{\nu} 
	+ \varepsilon^{\prime}\right)^2 - q^2} \ .
\label{eq.teta}
\end{equation}
The four-momentum transfer is $Q^{\mu} = k^\mu - k^{\prime\mu} = (\omega , \q)$, 
with $Q^2 = \omega^2 - q^2$. 
The factor ${\cal F}^2 $ comes from the contraction between the lepton and the 
hadron tensor and contains all the relevant nuclear structure information. 
${\cal F}^2 $ can be decomposed into a
charge-charge $(CC)$, charge-longitudinal $(CL)$, longitudinal-longitudinal
$(LL)$ and two transverse $(T ,T^{\prime})$ responses, i.e., 
\begin{eqnarray}	
{\cal F}^2_{+(-)} &=& V_{CC}R_{CC} 
+ 2 V_{CL}R_{CL} + V_{LL}R_{LL} +  V_{T}R_{T} \nonumber \\
&+& 2\chi V_{T^{\prime}}R_{T^{\prime}} \ , \label{eq.F}
\end{eqnarray}
where $\chi =+1 (-1)$ in the case of neutrino (antineutrino) scattering.
The coefficients $V$ are given by
\begin{eqnarray}
	V_{CC} &=& 1 - \delta ^2  \tan^2\left(\overline{\vartheta}/2\right)  \ , \ \ 
	V_{CL} = \frac{\omega}{q} + \frac{\delta ^2}{\rho^{\prime}}  
	\tan^2\left(\overline{\vartheta}/2\right)  \ , \nonumber \\
	V_{LL} &=& \left(\frac{\omega}{q}\right)^2 + 
	\left( 1 + \frac{2\omega}{q\rho^{\prime}} + \delta^2 \rho\right)  
	\delta^2\tan^2\left(\overline{\vartheta}/2\right) \ , \nonumber \\
	V_{T} &=&  \tan^2\left(\overline{\vartheta}/2\right)  + 
	\frac{\rho}{2} - \frac{\delta^2}{\rho^{\prime}} 
	\left(\frac{\omega}{q} + \frac{\rho\rho^{\prime}\delta^2}{2}\right)  
	\tan^2\left(\overline{\vartheta}/2\right) 
 \ , \nonumber \\
V_{T^{\prime}} &=& \frac{1}{\rho^{\prime}}
\left( 1 - \frac{\omega\rho^{\prime}\delta^2}{q}\right)
\tan^2\left(\overline{\vartheta}/2\right) \ , \label{eq.v}
\end{eqnarray}
where 
\begin{eqnarray}
	\delta = \frac{m}{\sqrt{|Q^2|}} \ , \ \ \ 
	\rho = \frac{|Q^2|}{q^2} \ , \ \ \ \rho^{\prime} = 
	\frac{q}{E_{\nu} + \varepsilon^{\prime}} \ . \label{eq.v1}
\end{eqnarray}
The weak response functions are given by
\inieq
R_{CC} &=& W^{00}\ ,\  R_{CL} = -(W^{03} +  W^{30})/2 \ ,\ R_{LL} = W^{33}\ , 
\nonumber \\ 
R_{T} &=& W^{11} +  W^{22} \ , \ 
R_{T^{\prime}} = -i(W^{12} -  W^{21})/2\ ,
\label{eq.response}
\fineq
in terms of the components of the hadron tensor,
that is given by suitable bilinear products of the transition matrix
elements of the nuclear many-body charged-current operator $\hat{J}^{\mu}$ between
the initial state $\mid\Psi_0\rangle$ of the nucleus, of energy $E_0$, and the 
final states $\mid \Psi_{{f}} \rangle$, of energy $E_{{f}}$, 
both eigenstates of the $(A+1)$-body Hamiltonian $H$, as 
\begin{eqnarray}
 W^{\mu\nu}(q,\omega) &=& \overline{\sum}_i
 \bint\sum_{ {f}}  \langle 
\Psi_{ {f}}\mid \hat{J}^{\mu}(\q) \mid \Psi_0\rangle \nonumber \\ &\times&
\langle 
\Psi_0\mid \hat{J}^{\nu\dagger}(\q) \mid \Psi_{ {f}}\rangle 
\ \delta (E_0 +\omega - E_{ {f}}),
\label{eq.ha1}
\end{eqnarray}
involving an average over the initial states and a sum over the undetected final 
states. The sum runs over the scattering states corresponding to all of the 
allowed asymptotic configurations and includes possible discrete states.  

In the QE region the scattering is described in the one-boson exchange approximation and 
in the relativistic impulse approximation (RIA) by the sum of incoherent processes 
involving only one nucleon scattering, i.e., the scattering occurs with only one 
nucleon, which is subsequently emitted, while the remaining nucleons of the target 
behave as spectators. Within the RIA the nuclear current operator is assumed to be the 
sum of single-nucleon currents $j^{\mu}$, corresponding to the weak CC operator 
\begin{eqnarray}
  j_{}^{\mu} &=& \big[F_1^{ V}(Q^2) \gamma ^{\mu} + 
             i\frac {\kappa}{2m_N} F_2^{ V}(Q^2)\sigma^{\mu\nu}Q_{\nu} 
	     \nonumber \\ 
	     &-&G_{ A}(Q^2)\gamma ^{\mu}\gamma ^{5} +
	     F_{ P}(Q^2)Q^{\mu}\gamma ^{5} \big] \tau^{\pm},
	     \label{eq.cc}
\end{eqnarray}
where $\tau^{\pm}$ are isospin operators, $\kappa$ is the anomalous part of the 
magnetic moment, 
and $\sigma^{\mu\nu}=\left(i/2\right)\left[\gamma^{\mu},\gamma^{\nu}\right]$.
$F_1^{ V}$ and $F_2^{ V}$ are the isovector Dirac and Pauli 
nucleon form factors, which are related to the corresponding electromagnetic form 
factor by the conservation of the vector current.  
$G_{ A}$ and $F_{ P}$ are the axial and 
induced pseudoscalar form factors which are usually parametrized as
\begin{eqnarray}
G_{ A} &=& \frac{g_{ A} }
        {\left(1+Q^2/M^2_{ A}\right)^2} \ , \\ 
F_{ P}&=& \frac{2m_NG_{ A}}{m^2_{\pi} + Q^2} \ , \label{eq.formf}
\end{eqnarray}
where $g_{ A} \simeq 1.267$, $m_{\pi}$ is the pion mass, $m_N$ the nucleon mass,
and $M_{ A} \simeq 1.03$ GeV is the axial mass~\cite{gal71,ABM,Bern02}.

Within the RIA framework and under the assumption of a shell-model description for nuclear 
structure,
the components of the hadron tensor are obtained from the sum, over all the 
single-particle (s.p.) shell-model states, of the squared absolute value of the 
transition matrix elements of the single-nucleon current
\inieq
\langle\chi_{{E}}^{(-)}(E)\mid  \hat{\jmath}^{\mu}(\q)\mid \varphi_n \rangle  \ ,
	     \label{eq.dko}
\fineq
where $\chi_{{E}}^{(-)}(E)$ is the scattering state of the emitted nucleon and the 
overlap $\varphi_n$ between the ground state of the target $\mid\Psi_0\rangle$ and the 
final state $ \mid n\rangle$ of the residual nucleus is a s.p. shell-model state.
In this work the bound nucleon states $\varphi_n$ are taken as 
self-consistent Dirac-Hartree solutions derived within a RMF
approach using a Lagrangian containing $\sigma$, $\omega$ and $\rho$ 
mesons~\cite{boundwf,Serot,adfx,lala,sha}. 

Different prescriptions are used to calculate the relativistic s.p. scattering wave 
functions. In the relativistic plane wave impulse approximation (RPWIA), FSI between the 
outgoing nucleon and the residual nucleus are neglected. In the relativistic 
distorted wave impulse 
approximation (RDWIA) FSI effects are accounted for by solving the Dirac equation with 
strong relativistic scalar and vector optical potentials. 
This approach has been very successfull in describing exclusive $(e,e^{\prime}p)$ 
data~\cite{Ud1,Ud3,Ud4,meucci1,meucci2,Kel,book}, but it would be 
inconsistent for the inclusive scattering, where 
all the inelastic channels should be retained and the total flux, although
redistributed among all possible channels due to FSI,
must be conserved. In the RDWIA (with complex potentials) the flux is not conserved and 
the inclusive $(e,e^{\prime})$ cross section is underestimated~\cite{ee,Chiara03,Jin,Kim}.
A simple way of estimating the inclusive response within the RIA is to use purely real 
potentials. In a first approach, the imaginary part of the phenomenological relativistic 
energy-dependent optical potentials~\cite{chc} is set to zero, thus retaining in the 
calculations only the real part. In a second approach, the 
scattering states are described by using the same real scalar and vector
RMF potentials considered in the 
description of the initial bound state $\varphi_n$. We refer to these two different FSI 
descriptions as real relativistic optical potential (rROP) and RMF calculations, 
respectively.

We note that the rROP conserves the flux and thus it is inconsistent with the exclusive 
process, where the final state includes only the elastic rescattering of the knocked-out 
nucleon and thus a complex optical potential adjusted to elastic proton scattering data 
must be used. Moreover,
the use of a real optical potential is unsatisfactory from a theoretical point of 
view, since it is an energy-dependent potential, reflecting the different contribution of 
open inelastic channels for each energy. Dispersion relation then dictates that the 
potential should have a nonzero imaginary term~\cite{hori}. On the other hand, the RMF 
model is based on the use of the same strong energy-independent real 
potential for both bound and scattering states, so that it fulfills the dispersion 
relation~\cite{hori} and, further, it maintains the continuity equation.

In a different approach, Green's function techniques may be 
exploited to derive the inclusive response. This formalism allows to reconstruct the flux 
into non-elastic channels in the case of inclusive scattering, starting from the complex 
(scalar and vector)
optical potential which describes elastic scattering data. The components of the hadron tensor are 
written in terms of the s.p. optical model Green's function. This is the result of 
suitable approximations, such as the assumption of a one-body current and subtler 
approximations related to the impulse approximation. The explicit calculation of 
the s.p. Green's function is avoided by using its spectral representation, which 
is based on a biorthogonal expansion in terms of a non Hermitian optical potential 
$\cal H$ and of its Hermitian conjugate $\cal H^{\dagger}$. Calculations require matrix 
elements of the same type as the RDWIA ones in Eq.~(\ref{eq.dko}), but involve 
eigenfunctions of both $\cal H$ and $\cal H^{\dagger}$, where the different sign of the 
imaginary part gives absorption in one case and gain of flux in the other case. 
Thus, in the sum over $n$ the total flux is redistributed and conserved.  
This relativistic Green's function model allows for a consistent treatment of 
FSI in the exclusive and in the inclusive scattering and gives a good 
description of $(e,e')$ data~\cite{ee,confee}. Detailed discussions of the RGF 
model can be found in Refs.~\cite{ee,cc,eenr,eeann,confee}.

\subsection{Scaling at the quasielastic peak}
\label{sec.sca}

Scaling arguments applied to neutrino reactions were presented for the first time in 
Ref.~\cite{amaro05}, where a phenomenological \lq\lq SuperScaling\rq\rq\  approach, denoted as 
SuSA, was proposed based on the assumed universality of the scaling
function for electromagnetic and weak interactions. Hence, the phenomenological scaling 
function extracted from the analysis of electron scattering data was used in order to 
make ``model-independent'' predictions for neutrino-nucleus cross
sections. The universal scaling behaviour is highly related to 
the single-boson exchange approximation and this justifies that both neutrino and 
electron 
scattering in QE conditions exhibit to a large extent similar scaling. A different way to 
analyze the problem consists in making use of a specific model that works properly in the 
description of electron scattering data, and apply it to neutrino reactions. In this 
case, although cross sections are \lq\lq model-dependent'', theoretical scaling functions 
corresponding to neutrinos can be derived and compared directly with the theoretical 
functions for electrons (obtained in the same model) and also with the experimental 
scaling function. Thus, the universality property can be tested within the framework of 
a specific model.

In this work we apply scaling arguments to the RMF and RGF models. Results are shown in 
the next section. The usual procedure to get the scaling function 
has been considered, i.e., dividing the inclusive
cross sections (\ref{eq.sigma}) by the appropriate single-nucleon charged-current 
$\nu N$ elastic cross section weighted by the corresponding proton/neutron number 
involved in the process. The scaling function $f(\psi')$ obtained with this
procedure depends on the dimensionless scaling variable $\psi'(q,\omega)$  
extracted from the RFG analysis that incorporates the typical momentum scale 
for the selected nucleus~\cite{mai1,cab1}.
The explicit expressions for the CC single-nucleon responses 
have already been presented in Refs.~\cite{amaro05,neutrino2}. In these single-nucleon 
responses, effects coming from the motion of the nucleons in the nucleus have been 
considered only up to first order in the
Fermi momentum $\eta_F\equiv k_F/m_N$. The use of the full results for the non-Pauli 
blocked regime, as shown in \cite{neutrino2}, leads to very minor differences without 
modifying the discussion and general conclusions presented in the work.

\begin{figure}[h]
\begin{center}
\includegraphics[width=9cm,height=9cm]{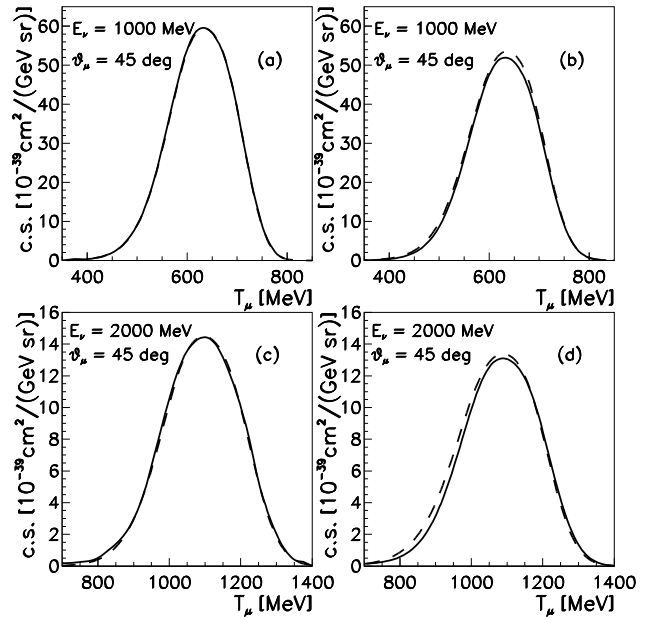} 
\end{center}
	\caption{Differential cross section of the $^{12}$C$(\nu_{\mu} , \mu ^-)$ reaction 
	as a function of the kinetic energy of the outgoing muon $T_\mu$ for two 
	values of the incident neutrino energy, i.e., E$_{\nu}$ = 1000 and 2000 
	MeV, and fixed muon 
	scattering angle $\vartheta_{\mu}$ = 45 deg, calculated by the Pavia 
	(solid lines) and the Madrid-Sevilla (dashed lines) groups with RPWIA 
	[(a) and (c)] and rROP [(b) and (d)].  \label{f1}}
	\end{figure}
\begin{figure}[h]
\begin{center}
\includegraphics[width=9cm,height=9cm]{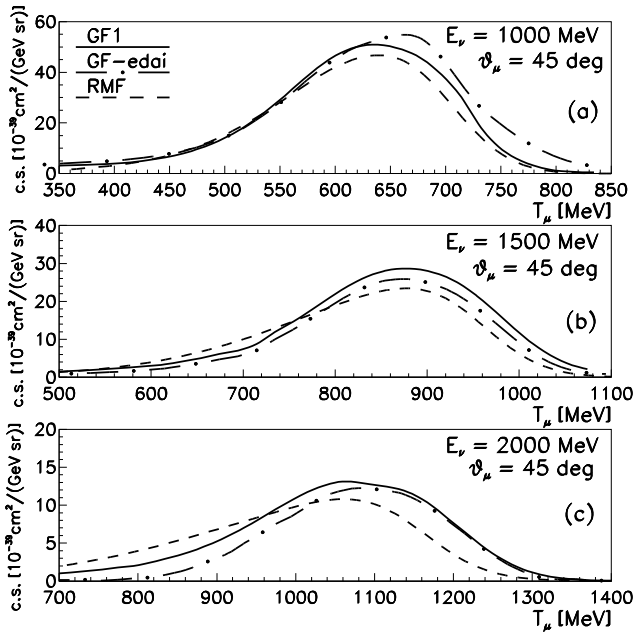} 
\end{center}
	\caption{Differential cross section of the $^{12}$C$(\nu_{\mu} , \mu ^-)$ 
	reaction for three 
	values of the incident neutrino energy, i.e., $E_{\nu}$ = 1000 (a), 
	1500 (b), 
	and 2000 MeV (c), and $\vartheta_{\mu}$ = 45 deg. The solid and long 
	dot-dashed lines are the RGF results calculated with the two different
        optical potentials 
	EDAD1 and EDAI-12C. The dashed lines are the results of the RMF model.
  \label{f2}}
	\end{figure}
\begin{figure}[h]
\begin{center}
\includegraphics[width=9cm,height=9cm]{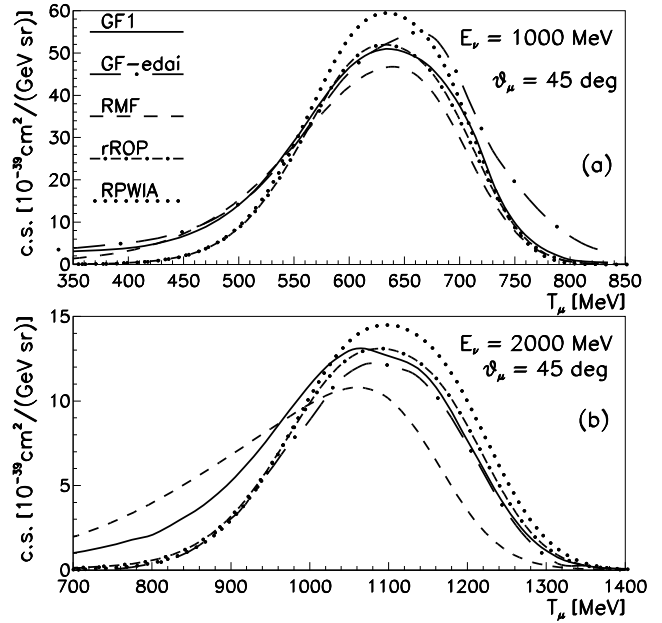} 
\end{center}
	\caption{Differential cross section of the $^{12}$C$(\nu_{\mu} , \mu ^-)$ reaction 
	for E$_{\nu}$ = 1000 (a) and 2000 MeV (b) and scattering 
	angle $\vartheta_{\mu}$ = 45 deg.  The solid, long dot-dashed, and dashed 
	lines are the same as in  Fig. \ref{f2}. The dot-dashed and dotted lines are 
	the rROP and the RPWIA results, respectively. \label{f3}}
	\end{figure}
\begin{figure}[h]
\begin{center}
\includegraphics[width=9cm,height=9cm]{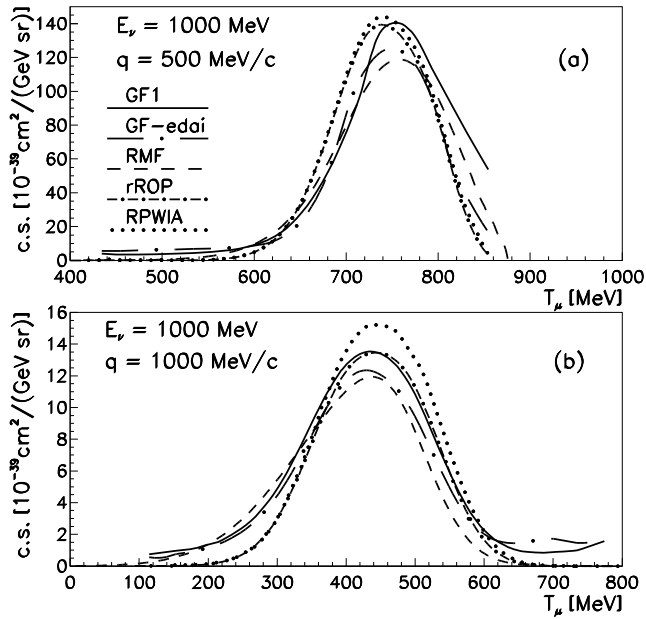} 
\end{center}
	\caption{Differential cross section of the $^{12}$C$(\nu_{\mu} , \mu ^-)$ reaction 
	for E$_{\nu}$ = 1000 MeV and $q$ = 500 MeV/$c$ (a) and 
	1000 MeV/$c$ (b). 
	Line convention as in Fig. \ref{f3}.  \label{f4}}
	\end{figure}
\begin{figure}[h]
\begin{center}
\includegraphics[width=9cm,height=9cm]{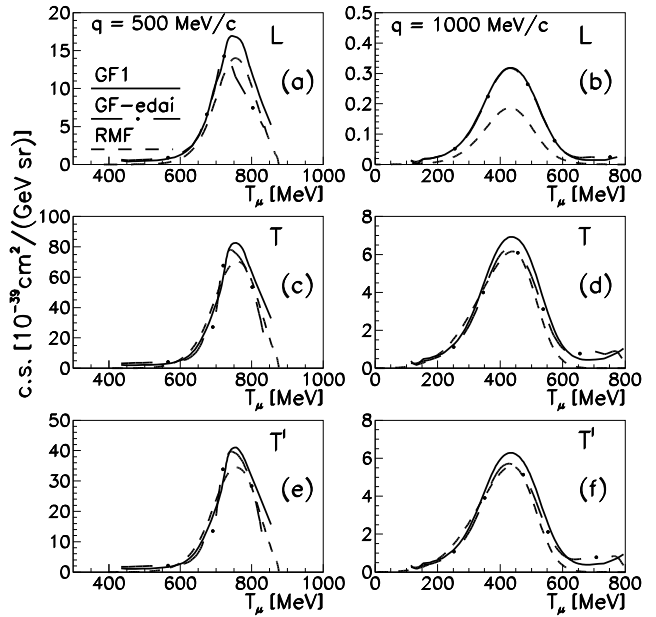} 
\end{center}
	\caption{$L$, $T$, and $T'$ components of the cross section 
	of the $^{12}$C$(\nu_{\mu} , \mu ^-)$ reaction for E$_{\nu}$ = 1000 MeV and 
	$q$ = 500 MeV/$c$ (panels (a), (c), and (d) in the left column) and 
	1000 MeV/$c$ (panels (b), (d), and (e) in the right column).  
	Line convention as in Fig. \ref{f2}.  \label{f5}
}
	\end{figure}
\begin{figure}[h]
\begin{center}
\includegraphics[width=9cm,height=9cm]{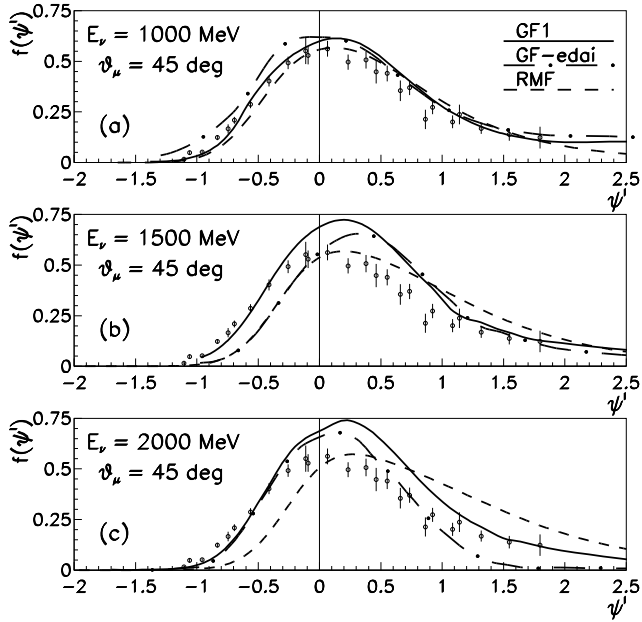} 
\end{center}
	\caption{Scaling function of the $^{12}$C$(\nu_{\mu} , \mu ^-)$ reaction in the 
	same kinematics as in Fig. \ref{f2} compared with the averaged experimental 
	scaling function. 
	Line convention as in Fig. \ref{f2}.  \label{f6}}
	\end{figure}
\begin{figure}[h]
\begin{center}
\includegraphics[width=9cm,height=9cm]{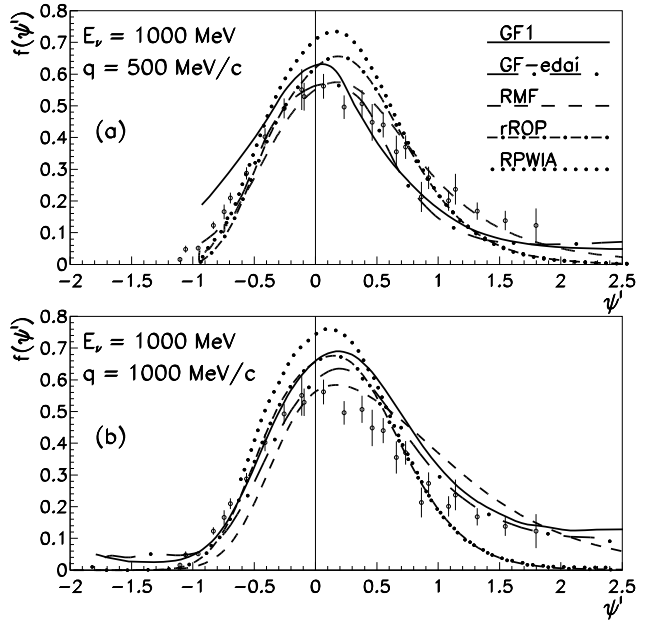} 
\end{center}
	\caption{Scaling function of the $^{12}$C$(\nu_{\mu} , \mu ^-)$ reaction 
	for incident neutrino 
	energy E$_{\nu}$ = 1000 MeV and $q$ = 500 MeV/$c$ (a) and 
	1000 MeV/$c$ (b). 
	Line convention as in Fig. \ref{f4}.  \label{f7}}
	\end{figure}
\begin{figure}[h]
\begin{center}
\includegraphics[width=9cm,height=9cm]{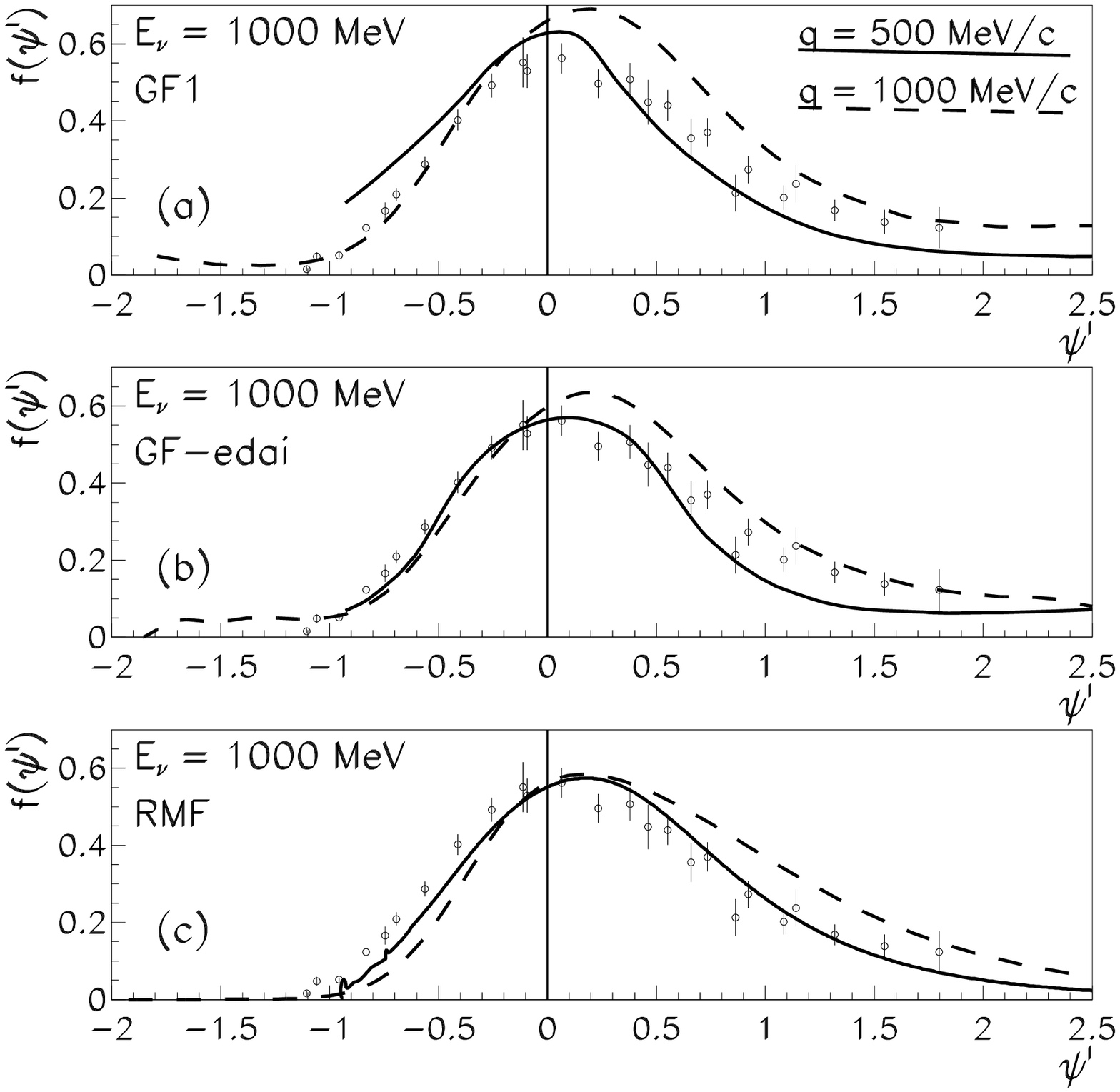} 
\end{center}
	\caption{Analysis of first kind scaling, $f({\psi'})$ for  
	E$_{\nu}$ = 1000 MeV and $q$ = 500 MeV/$c$ (solid lines) and 
	1000 MeV/$c$ (dotted lines) with the GF1 (a), GF-edai (b), and RMF (c) 
	models using the same results already displayed in Fig. \ref{f7}.  
	\label{f8}}
	\end{figure}

\section{Results and discussion}
\label{results}

In this section the numerical results of the different relativistic models developed by 
the Pavia and the Madrid-Sevilla groups to describe FSI in the inclusive
quasielastic CC neutrino-nucleus scattering are considered. As a first step results 
obtained by the two groups in the RPWIA and rROP approaches are compared to check the 
consistency of the numerical programs when calculations are carried out under the same 
conditions. Then the results corresponding to the RMF model developed by the 
Madrid-Sevilla group and the 
RGF model developed by the Pavia group are compared. The comparison is 
performed for the 
$^{12}$C$(\nu_{\mu},\mu^{-})$ cross section and scaling function calculated in different 
kinematics which are representative of the kinematical range where QE scattering is 
expected to give the main contribution to the inclusive neutrino-nucleus scattering. 
Pion-production events may play a significant role and increase their
contribution with larger values of the energy transfer. However,
its contribution for the CCQE MiniBone data is expected to be very minor: 
no pions in the final state are detected.
Antineutrino-induced scattering can also be calculated, but it is not considered in this 
paper. All the calculations are performed using the same relativistic initial state 
wave functions, that are taken as Dirac-Hartree 
solutions of a relativistic Lagrangian written in the context of a relativistic mean field 
theory with the NLSH parameterization~\cite{adfx,lala,sha}.

\subsection{Differential cross section}
  
The $^{12}$C $(\nu_{\mu},\mu^{-})$ cross section calculated by the Pavia and the Madrid-Sevilla 
groups in RPWIA and in rROP for two values of the incident neutrino energy, i.e., 
$E_{\nu} = 1000$ and $2000$ MeV, and a fixed muon scattering angle $\vartheta_{\mu} = 45$ 
degrees are compared in Fig.~\ref{f1}. 
Almost identical results are obtained in RPWIA. In rROP the two results are very similar, 
up to a few percent. A similar agreement between the RPWIA and rROP results of the Pavia 
and
Madrid-Sevilla groups was already found in Ref.~\cite{confee} for the inclusive QE 
electron scattering. 
The comparison in Fig.~\ref{f1} is an important and necessary 
benchmark of our independent computer programs, which 
allows us to estimate the numerical uncertainties and 
gives enough confidence on the reliability of both calculations. 

Cross sections evaluated with the RMF and RGF models for the  
kinematics with  $E_{\nu}$ = 1000, 1500, and 2000 MeV and 
$\vartheta_{\mu} = 45$ degrees are presented in Fig.~\ref{f2}.
In the case of the RGF approach two different parameterizations for the relativistic 
optical potential have been used: the energy-dependent and A-dependent EDAD1 and the 
energy-dependent but A-independent EDAI-12C complex phenomenological potentials 
of~\cite{chc}, which are fitted to proton elastic scattering data on several nuclei in an 
energy range up to 
1040 MeV. In the figures the results obtained with EDAD1 and EDAI-12C are denoted as GF1 
and GF-edai, respectively. We note that all the RGF results  presented here 
contain the contribution of both terms of the hadron tensor in Eq.~(25) of 
Ref.~\cite{cc}, while only the contribution of the first term was included in 
the numerical results presented in Ref.~\cite{cc}. The second term is entirely 
due to the imaginary part of the optical potential and vanishes if only the 
real part is considered. The calculation of this term is a complicated and time
consuming task that requires integration over all the eigenfunctions of the 
continuum spectrum of the optical potential. 
Indeed the first term gives in general the main contribution.  
The contribution of the second term depends on kinematics and becomes more 
important for higher values of the energy transfer, i.e., higher energies for 
the optical potential. Although small and even negligible in many situations,
the second term can be very important in particular kinematic conditions and 
in general cannot be neglected. 

To make the comparison of the different models clearer, we compare in Fig.~\ref{f3} the 
RMF and RGF results directly with the RPWIA and rROP ones. 
We note that in these kinematics, that are similar to those actually explored at present 
neutrino experiment facilities, the momentum transfer $q$ is not fixed and its value 
around the peak of the cross section is usually large, i.e., $q \approx 700$ MeV$/c$ for 
$E_{\nu} = 1000$ MeV, $q \approx 1100$ MeV$/c$ for $E_{\nu} = 1500$ MeV, and 
$q \approx 1400$ MeV$/c$ for $E_{\nu} = 2000$ MeV.
The shape of the RMF cross section shows an asymmetry, with a long tail extending towards 
lower values of the energy of the final muon, i.e., higher values of $\omega$, which is 
due to the strong energy-independent scalar and vector potentials present in the RMF 
approach. The asymmetry increases with larger
incident neutrino energies. In the case of the RGF cross sections the 
asymmetry towards higher $\omega$ and $E_{\nu}$ is less significant but still visible. 
Almost no asymmetry is found for the RPWIA and rROP cross sections. We note that in the present 
calculations, that are performed in kinematics different from those considered in 
Ref.~\cite{cc}, significant
differences  are obtained between the rROP and RGF results. These differences stress the 
important role played by the imaginary part of the optical potential, which in the RGF 
model affects both terms of the hadron tensor. 
The GF1 and GF-edai cross sections have somewhat different shapes, that are particularly 
visible for low $\omega$ at  $E_{\nu} = 1000$ MeV and for higher 
$\omega$ at $E_{\nu} = 2000$ MeV. These differences are essentially due
to the imaginary part of the ROP, that is sensitive to the particular parametrization 
considered. The real terms of the phenomenological ROP are 
very similar for different parametrizations and give very 
similar results. 

The differences between the results of the GF1 and GF-edai cross sections depend on the 
energy and momentum transfer, and are directly linked to the specific structure of the 
energy-dependent relativistic optical potentials adopted in the RGF model. The differences 
between the RMF and RGF results increase with the energy and momentum transfer. The larger 
differences seen for the largest value of $E_{\nu}$, not only between the RMF and RGF 
models, but also between the two RGF results, is simply an indication of the difference in 
the ingredients of these calculations.

To perform a more direct comparison between our neutrino- and 
electron-scattering calculations of Ref.~\cite{confee}, we present in Fig.~\ref{f4} our 
cross sections calculated for a fixed value of the incident 
neutrino energy, $E_{\nu} = 1000$ MeV,  and two values of the momentum 
transfer, i.e., $q$ = 500 and 1000 MeV$/c$. Also in these kinematics the shape of the RMF 
cross section shows an asymmetry with a tail extending towards higher values of $\omega$. 
An asymmetric shape 
towards higher $\omega$ is shown also by the RGF cross sections, while no visible asymmetry is 
given by the RPWIA and rROP results. Also in these kinematics significant differences are 
obtained between the RGF and rROP cross sections.

The behavior of the RMF and RGF results as a function of $q$ and $\omega$ is connected 
to the different relativistic potentials used in the models. 
The RMF model uses
the effective mean field, that reproduces the saturation behavior of nuclear
matter and the properties of the ground state of nuclei.
On the other hand, the RGF model uses the phenomenological optical potentials fitted to
elastic proton-nucleus scattering. The lost of 
elastic flux into inelastic channels caused by the imaginary 
term of these potentials is recovered for inclusive scattering in the RGF making use
of dispersion relations. As already shown for $(e,e')$ reactions~\cite{confee},
the RGF yields a larger cross section than the RMF. The latter can be considered
as an estimation of the the pure nucleonic contribution. On the contrary, the RGF 
may account for in a phenomenological way additional effects due to non nucleonic 
degrees of freedom, more important with increasing transferred momentum. 
Notice that both the RGF and the RMF yield predictions within 
few percent for low-$q$ (panels (a) of Figs.~2 and 3), while
as $q$ goes up the RGF yields increasingly larger cross sections than RMF. This
may reflect the influence of the pionic degrees of freedom.

The results obtained with GF1 and GF-edai are consistent with the general 
behavior of the phenomenological energy-dependent relativistic optical 
potentials and the differences are basically due to their different imaginary 
part, that includes the overall effect of the inelastic channels and is not
univocally determined by the elastic phenomenology.

The results displayed in Fig.~\ref{f4} present some differences with respect to the 
corresponding cross sections of the inclusive electron scattering shown in 
Ref.~\cite{confee}, which are calculated with the same 
models for FSI and at the same values of the momentum transfer. In both cases the 
differences
between the results of the different models are generally larger for increasing value of 
the momentum transfer. For neutrino scattering, however, such a behavior is less evident 
and clear.  In particular, the GF1 cross section at $q$ = 1000 Mev$/c$ does not show the 
strong enhancement in the region of the maximum that was found for the $(e,e^{\prime})$ 
calculations of Ref.~\cite{confee}, where the GF1 result 
was even larger than the RPWIA one. In the case of neutrino scattering the RGF results in 
the region of the maximum are
generally larger than the RMF ones, but smaller than the RPWIA cross sections. 

In spite of many similarities, inclusive electron 
scattering and CC neutrino-nucleus scattering are two different processes and 
caution should be drawn on their comparison.
We note that the cross sections shown in Fig.~\ref{f4} are calculated with same incident 
lepton energy and the same momentum transfer as in the $(e,e^{\prime})$ 
calculations of Ref.~\cite{confee}. Nevertheless the muon mass gives in the 
case of 
CC neutrino scattering a different kinematics, with different values of the energy 
transfer and, as a consequence, of the energies of the outoing nucleon. 
This means that in the RGF model the optical potential is calculated for 
electron and neutrino scattering at different energies. To clarify this
point and reproduce the kinematics of electron scattering, we have performed 
some calculations for the $(\nu , \mu ^-)$ reaction with vanishing muon mass. 
The main difference with respect to the calculations shown in Fig.~\ref{f4} is 
a shift of the cross section, by about 100 MeV, towards higher values of 
$T_{\mu}$, without any significant change in the shape or in the strength. 

The most important difference between electron- and neutrino-scattering 
at the Feynman amplitude level is the exchanged vector boson: a virtual photon 
probing the electromagnetic current in electron scattering, and a $W^{\pm}$ 
probing the weak current in CC neutrino-nucleus scattering. Thus, whereas for 
the $(e,e^{\prime})$ process all nucleons (protons and neutrons) in the nucleus
contribute, in the case of CC neutrino (antineutrino) scattering only the neutrons 
(protons) in the target contribute to the inclusive cross section. 
Moreover, whereas the $R_L$ and $R_T$ responses in CC neutrino induced scattering
are given by the sum of pure vector-vector (VV) and axial-axial (AA) terms
(see, e.g. Refs.~\cite{amaro05,neutrino2}), the
$R_{T^\prime}$ response comes from the interference between the vector (V) and axial (A)
terms in the neutrino CC operator.
In the electron scattering case, where both isoscalar and isovector 
form factors enter the responses, the cross section is the sum of the longitudinal and 
transverse responses. 

The different currents and their possible interplay with the other 
ingredients of the models do not allow an easy comparison between 
the results of electron and neutrino scattering.
The numerical differences between the RGF results for electron and neutrino 
scattering can mainly be ascribed to the combined effects of the weak current, 
in particular, its axial term, and the imaginary part of the relativistic 
optical potential. We have checked that these effects can explain 
the fact that in neutrino scattering the GF1 result does not give the strong 
enhancement in the region of the maximum of the cross section that was found for 
the $(e,e^{\prime})$ calculations of Ref.~\cite{confee}.

As a further example, we compare in Fig.~\ref{f5} the RMF and RGF results for 
the longitudinal, transverse, and interference axial-vector
 contributions to the cross section. The sum of these three 
terms gives the corresponding cross sections in Fig.~\ref{f4}. Here, by 
longitudinal contribution to the cross section we mean the sum of the $R_{LL}$, 
$R_{CL}$, and $R_{CC}$ responses, defined in Eq.~\ref{eq.response}, each one 
multiplied by the appropriate kinematical factors of Eq.~\ref{eq.v}.
In general, the RMF and RGF results as a function of $q$ and $\omega$ show the 
same behavior as the cross sections in Fig.~\ref{f4} and, again, are related to 
the different relativistic potentials used in the models. The $T$  and 
${T^{\prime}}$ components are similar for all cases: at $q = 500$ MeV$/c$  
both RGF results are larger than the RMF ones in the peak region, while at 
$q = 1000$ MeV$/c$ the RMF and GF-edai results are very similar and lower than 
GF1. The longitudinal cross section has a different behavior: at 
$q = 500$ MeV$/c$ the GF-edai result is close to RMF and lower than GF1, while 
at $q = 1000$ MeV$/c$ it overlaps GF1 and they both are larger than RMF.
At $q = 500$ MeV$/c$ all the GF and RMF calculations give the $T$ cross section
approximately two times larger than the ${T^{\prime}}$ one. In addition, the 
longitudinal contribution, although smaller, cannot be neglected; in this case the relative 
difference between the GF1 and GF-edai longitudinal components is large and 
has visible effects in the cross sections of Fig.~\ref{f4}.
At $q = 1000$ MeV$/c$ the $T$ and ${T^{\prime}}$ components have similar 
strengths, whereas the longitudinal response gives only a very small 
contribution to the cross section. 
In the electron scattering case, at $q = 500$ MeV$/c$ the longitudinal and 
transverse components have similar strength whereas at $q = 1000$ MeV$/c$ the 
transverse response is much more important (see, e.g., Refs.~\cite{ee,confee}).

\subsection{Scaling functions}

The effects already discussed for the differential cross sections are also 
present in the scaling functions. Here we compare results for the scaling 
function extracted from the differential cross section
$f_{}(\psi^{\prime})$  using the 
same descriptions for the final state interactions already considered 
for the  cross sections. The
transverse responses $R_{T}$ and $R_{T^\prime}$ are the leading terms for
the total cross section whereas the longitudinal contribution
$R_{L}$ is usually small; the corresponding scaling function 
$f_{T}(\psi^{\prime})$ and $f_{T^\prime}(\psi^{\prime})$ are very similar, 
up to a few percent, and, moreover, they are similar to the scaling function 
$f_{}(\psi^{\prime})$ from the cross section, i.e., scaling
of zero kind is verified. Thus, we do not show $f_{T}(\psi^{\prime})$ and 
$f_{T^\prime}(\psi^{\prime})$  and present our results only for 
$f_{}(\psi^{\prime})$. The longitudinal contribution leads
to a scaling function $f_{L}(\psi^{\prime})$ which may depart significantly
from $f_{}(\psi^{\prime})$. However, one should be cautious because the
longitudinal contribution to inclusive neutrino-nucleus cross sections can be almost
negligible in some kinematics compared with the transverse $T$ and $T^{\prime}$ ones. 

As a first step, we have compared the scaling function $f_{}(\psi^{\prime})$  
obtained by the Pavia and Madrid-Sevilla groups in RPWIA and rROP for three 
values of the neutrino energy, $E_{\nu}$ = 1000, 1500, and 2000 MeV, and 
$\vartheta_{\mu} = 45$ degrees. Our results are almost coincident and, in 
addition to the cross section results in Fig.~\ref{f1}, confirm the
consistency of the numerical codes when calculations are performed under the
same conditions. We do not show these results for simplicity.

In Figs.~\ref{f6} and~\ref{f7} we compare the scaling function $f(\psi')$ 
evaluated with different models and for the same kinematical conditions as in 
Figs.~\ref{f2} and~\ref{f4}, respectively, with the averaged QE phenomenological 
longitudinal scaling function extracted from the analysis of $(e,e^{\prime})$ 
data~\cite{don1,don2,mai1}. 
As already shown in previous works~\cite{cab1,cab2,isospin07}, the RMF model 
produces an asymmetric shape, 
with a long tail in the region with $\psi^{\prime}>0$, that follows closely the 
behavior of the phenomenological 
function.  It has been noticed that
the RMF approach is capable of explaining the asymmetric behavior of the
phenomenolgical scaling function within 
the framework of the relativistic impulse approximation taking advantage of its 
strong relativistic scalar and vector potentials. 
The results with the RGF model in Fig.~\ref{f6} are similar to those obtained 
with RMF at $E_{\nu}$ = 1000 MeV, while visible discrepancies appear at 
$E_{\nu}$ = 1500 and 2000 MeV, and in Fig.~\ref{f7}. 
As a general remark, these results for the scaling functions follow similar
trends to those already applied to the behavior of the cross sections in 
Figs.~\ref{f2} and \ref{f4}, i.e., the differences between the RMF and RGF, and 
between the GF1 and GF-edai calculations, increase with larger energy and momentum
transfer. 

The asymmetric shape with a tail in the region of  positive $\psi^{\prime}$ is 
obtained in both RMF and RGF models, which involve descriptions of 
FSI either with a strong energy-independent real potential or with a complex 
energy-dependent optical potential, respectively.
The scaling functions corresponding to RPWIA and rROP, which are also presented 
in Fig.~\ref{f7} for two values of the momentum transfer, do not show any 
significant asymmetric tail for $\psi^{\prime}>0$.

It has been already discussed~\cite{cab1,cab2,isospin07} that the RMF model 
produces a scaling function $f(\psi^{\prime})$ that is in accordance with the 
longitudinal scaling function extracted from 
the $(e,e^{\prime})$ reaction and hence with the electron scattering 
longitudinal data.
This outcome reinforces the validity of the general 
assumption~\cite{cab1} to predict CC neutrino-nucleus
cross sections from the phenomenological scaling function, and 
it is a noticeable fact that this is contrary to what one
would expect, since 
neutrino reactions are totally dominated by the transverse 
$R_{T}$ and $R_{T^{\prime}}$ responses. 

In the case of the RGF model we apparently do not obtain a similar good 
agreement with the longitudinal scaling function 
$f_{L}(\psi^{\prime})$ extracted from the $(e,e^{\prime})$ results of 
Ref.~\cite{confee}. 
This is clearly related to the fact that, at higher $q$
values, there is not such a strong enhancement of the maximum strength, 
particularly for GF1 as in Ref.~\cite{confee}. However, taking advantage of 
this, the RGF model produces a scaling function in reasonable agreement with 
the averaged experimental scaling function. 

An analysis of the scaling of first-kind, i.e., independence of the momentum 
transfer, is illustrated in Fig.~\ref{f8}. The results are the same already
shown in Fig.~\ref{f7}, but are presented in a different way. Each panel
corresponds to a specific description of FSI (GF1, GF-edai, and RMF)  
and includes the results obtained at $q$ = 500 and 1000 MeV$/c$.
The experimental $(e,e^{\prime})$ data are compatible with a mild violation of 
the first-kind scaling, particularly in the positive $\psi'$-region.
In Refs.~\cite{cab1,cab2} the scaling functions evaluated with the RPWIA and 
rROP models were shown to depend very mildly on the transferred momentum in the whole,
positive and negative, $\psi^{\prime}$ region.
In the case of the RMF approach, there is a slight shift in the region 
$\psi^{\prime}<0$ (slighter than for the longitudinal scaling function of
electron scattering), whereas the model breaks scaling approximately at $30\%$ 
level when $\psi^{\prime}>0$. Similar results are obtained with the 
RGF models, where a small shift in the region of negative $\psi^{\prime}$ also occurs,
and scaling is broken for $\psi^{\prime}>0$. This scaling violation for 
$\psi'>0$ is slightly larger with GF1 than GF-edai. However, the quality of the 
comparison between the experimental QE scaling function
and the two RGF results is similar. Finally, the stronger scaling breakdown shown by RGF,
compared to RMF, may reflect
the effective presence of non-nuclenic contributions in the case of RGF.


\section{Summary and conclusions}
\label{conc}

This work emerges from the collaboration between the Pavia and Madrid-Sevilla groups,
and it extends the study presented in Ref.~\cite{confee} where we
focused on inclusive electron scattering reactions. Here we consider the case of 
charged-current quasielastic neutrino-nucleus scattering processes. These studies
are of great interest in connection with the recent data measured at 
MiniBooNE~\cite{miniboone},
and their \lq\lq possible\rq\rq \ impact on neutrino oscillations. 

The kinematics of the MiniBooNE experiment involves neutrino energies up to $\sim 3$ 
GeV. This implies that relativity is an essential ingredient; not only relativistic 
kinematics
should be considered, but also nuclear dynamics and current operators should be described 
within a relativistic framework. The analysis performed by our groups involves
a relativistic description of the reaction mechanism with strong scalar and vector
potentials, thus providing a proper 
treatment of the neutrino scattering process of interest. However, before entering into
a detailed analysis of the MiniBooNE experiment and its comparison with our theoretical 
results, that will be considered in a subsequent work, in this paper we restrict ourselves 
to a systematic comparison between the two approaches for different kinematics. As already
mentioned in Ref.~\cite{confee}, the description of the final state interactions constitutes a 
basic aspect in the analysis of the theoretical results and their comparison with
data.

Following Ref.~\cite{confee}, we check the consistency of the numerical calculations by 
comparing results in the plane wave limit (RPWIA) and also making use of the real part
of the relativistic optical potentials (rROP). Almost identical results are obtained in 
RPWIA and very close in the case of rROP. This outcome is consistent with our previous
analysis in electron scattering, and it reinforces the reliability of the calculations. 
Next we apply the two models considered in Ref.~\cite{confee} to account for FSI, that is,
the RMF, based on the use of the same relativistic mean field potential for the bound and 
ejected nucleon wave functions, and the RGF approach, which treats FSI consistently
in the inclusive and exclusive reactions. In the latter we have considered two different 
optical potentials, EDAD1 and EDAI-12C, which have been fitted to proton elastic scattering
data. 

Contrary to RPWIA and rROP approaches, the differential cross sections obtained within RMF 
and RGF show a significant asymmetry with a tail extended to large values of the energy 
transfer (small muon kinetic energies). The amount of this asymmetry in the RMF approach 
increases with larger neutrino energies. On the contrary, its contribution is less 
significant in the RGF model, showing also important differences attached to the two 
optical potentials considered. These discrepancies are linked to the imaginary 
energy-dependent term in the ROP that is very sensitive to the particular parameterization 
considered. It seems that optical potentials fitted only to elastic 
scattering data are not well constrained when applied, within the RGF approach, to 
describe inclusive reactions. In addition to these effects introduced by the particular 
ROP considered, also significant differences emerge from the comparison between RMF and 
RGF, being larger for increasing transferred momentum. 
Although this general behavior was already present in the case of $(e,e')$ 
processes~\cite{confee}, the results for neutrino reactions show some distinct 
features to be discussed. In particular, the GF1 $(\nu,\mu^-)$ cross section does 
not present the strong enhancement in the region of the maximum that was previously 
found for $(e,e')$ at $q=1000$ MeV/c. This result has implications for the scaling 
function. As shown in Ref.~\cite{isospin07}, the function $f(\psi')$ obtained for CCQE 
neutrino reactions within the RMF model is in accordance with $f_L(\psi')$ from 
$(e,e')$. This is connected to the isospin content in the nucleon form factors. 
On the contrary, such an agreement does not occur for RGF. This would indicate 
that the universality property of the superscaling function is not entirely 
fulfilled within the RGF model.

Summarizing, we present a systematic analysis of cross sections and scaling properties for
CCQE neutrino-nucleus scattering processes. We consider two different relativistic
descriptions of the final state interactions. Although both approaches respect
scaling and superscaling behavior, some differences emerge, particularly concerning the
comparison between the results obtained for electrons and neutrinos. The 
present results should be cautiously viewed due to the very different 
ingredients considered by RMF and RGF models; however, these may help in 
disentangling different physics aspects involved in the processes, with a 
special mention to the analysis of the MiniBooNE data and FSI effects. 
Work along this line is presently in progress.


\begin{acknowledgments}

We are grateful to M.B. Barbaro and F.D. Pacati for useful discussions and for their 
valuable advice. 
This work was partially supported by DGI (Spain) under contract 
nos. FIS2008-04189, FPA2010-17142, the Spanish Consolider-Ingenio 2010 programme CPAN 
(CSD2007-00042), and by the INFN-CICYT collaboration agreements FPA2008-03770-E and 
ACI2009-1053.

\end{acknowledgments}



\end{document}